\documentclass[mathleft
]{an}
\usepackage{graphicx}
\usepackage{times}
\usepackage{amssymb,amsmath}

\usepackage{natbib}
\begin{document}

\Pagespan{1}{}
\Yearpublication{2013}%
\Yearsubmission{2012}%
\Month{11}%
\Volume{999}%
\Issue{88}%

\title{Magnetic fields during high redshift structure formation}

\author{Dominik R. G. Schleicher\inst{1}\fnmsep\thanks{Corresponding author:
  \email{dschleic@astro.physik.uni-goettingen.de}\newline}
\and  Muhammad Latif\inst{1}
\and Jennifer Schober\inst{2}
\and Wolfram Schmidt\inst{1}
\and Stefano Bovino\inst{1}
\and Christoph Federrath\inst{3}
\and Jens Niemeyer\inst{1}
\and Robi Banerjee\inst{4}
\and Ralf S. Klessen\inst{2}
}
\titlerunning{Magnetic fields at high redshift}
\authorrunning{Schleicher et al.}
\institute{
Institut f\"ur Astrophysik, Georg-August-Universit\"at G\"ottingen, Friedrich-Hund-Platz 1, 37077 G\"ottingen, Germany
\and 
Universit\"at Heidelberg, Zentrum f\"ur Astronomie der Universit\"at Heidelberg, Institut f\"ur theoretische Astrophysik, Albert-\"Uberle-Str. 2, 69120 Heidelberg, Germany
\and 
Monash Centre for Astrophysics, School of Mathematical Sciences, Monash University, Vic 3800, Australia
\and
Universit\"at Hamburg, Hamburger Sternwarte, Gojenbergsweg 112, 21029 Hamburg
}

\received{30 May 2005}
\accepted{11 Nov 2005}


\abstract{%
  We explore the amplification of magnetic fields in the high-redshift Universe. For this purpose, we perform high-resolution cosmological simulations following the formation of primordial halos with $\sim10^7$~M$_\odot$, revealing the presence of turbulent structures and complex morphologies at resolutions of at least $32$ cells per Jeans length. Employing a turbulence subgrid-scale model, we quantify the amount of unresolved turbulence and show that the resulting turbulent viscosity has a significant impact on the gas morphology, suppressing the formation of low-mass clumps. We further demonstrate that such turbulence implies the efficient amplification of magnetic fields via the small-scale dynamo. We discuss the properties of the dynamo in the kinematic and non-linear regime, and explore the resulting magnetic field amplification during primordial star formation. We show that  field strengths of $\sim10^{-5}$~G can be expected at number densities of $\sim5$~cm$^{-3}$.
  }

\maketitle

\section{Introduction}

In the present-day Universe, magnetic fields  have been detected in galaxies \citep{Beck04}, galaxy clusters \citep{Kim90} and local dwarf galaxies \citep{Chyzy11, Heesen11, Kepley11}. Even for the intergalactic medium, more speculative observations suggest the presence of magnetic fields \citep{Neronov10, Tavecchio11, Broderick12, Miniati12, Takahashi12, Yuksel12}. Magnetic fields have further been inferred in high-redshift galaxies via the observed correlation of quasar rotation measures with the number of galaxies along the line-of-sight until $z\sim2$ \citep{Bernet08, Kronberg08}, as well as through the observed far-infrared - radio correlation, which is established until $z\sim4$ \citep{Murphy09}.

We propose here that strong magnetic fields are produced as a result of turbulent magnetic field amplification already at high redshift. The small-scale dynamo efficiently converts turbulent into magnetic energy, on timescales considerably smaller than the dynamical time \citep{Kazantsev68, Brandenburg05, Schober12c}. As a result, the dynamo process will likely produce strong magnetic fields during the formation of the first stars and galaxies \citep{Schleicher10c, Sur10, Federrath11, Sur12, Schober12, Turk12, Peters12}.

A central condition for this scenario is the presence of turbulence.  The latter has  been reported by \citet{Abel02} for the first star-forming minihalos, and \citet{Greif08} and \citet{Wise08} for larger primordial halos. However, such massive halos were only explored with a resolution of $\sim16$ cells per Jeans length, while \citet{Federrath11} have shown that at least $32$ cells per Jeans length are required in order to obtain converged turbulent energies. 

In this paper, we present a detailed study of the turbulent properties of massive primordial halos, exploring the effects of resolved turbulence by systematically increasing the numerical resolution per Jeans length, as well as the unresolved turbulence using the subgrid-scale turbulence model by \citet{Schmidt11}. We then discuss the amplification of magnetic fields in the kinematic and non-linear regime. Our results indicate the likely presence of strong magnetic fields in the high-redshift Universe.

\section{Turbulence during the formation of protogalaxies}

To explore the generation of turbulence during the formation of protogalaxies, we perform cosmological hydrodynamics simulations with the adaptive-mesh refinement (AMR) code Enzo\footnote{Enzo website: http://code.google.com/p/enzo/} \citep{OShea04}. We explore the evolution of a cosmological box of $1$~Mpc/h with a root grid resolution of $128^3$ and two initial nested $128^3$ grids centered on the most massive halo, each increasing the resolution by a factor of 2 \citep{Latif12b}. We include primordial chemistry following the non-equilibrium evolution of H, H$^+$, e$^-$, H$^-$, H$_2^+$ and H$_2$, along with a photodissociating background corresponding to $J_{21}=1000$, where $J_{21}$ is given in units of erg~s$^{-1}$~cm$^{-2}$~Hz$^{-1}$~sr$^{-1}$. Our version of the code includes a subgrid-scale model for unresolved turbulence \citep{Maier09, Schmidt11}, accounting for turbulent pressure, diffusion and including an explicit modeling of the turbulent energy dissipation. Our main refinement criterion is the number of cells per Jeans length, where we explore values of $16$, $32$ and $64$. We stop our simulations when the highest refinement level $27$ is reached.

\begin{figure*}
\begin{center}
\includegraphics[scale=0.6]{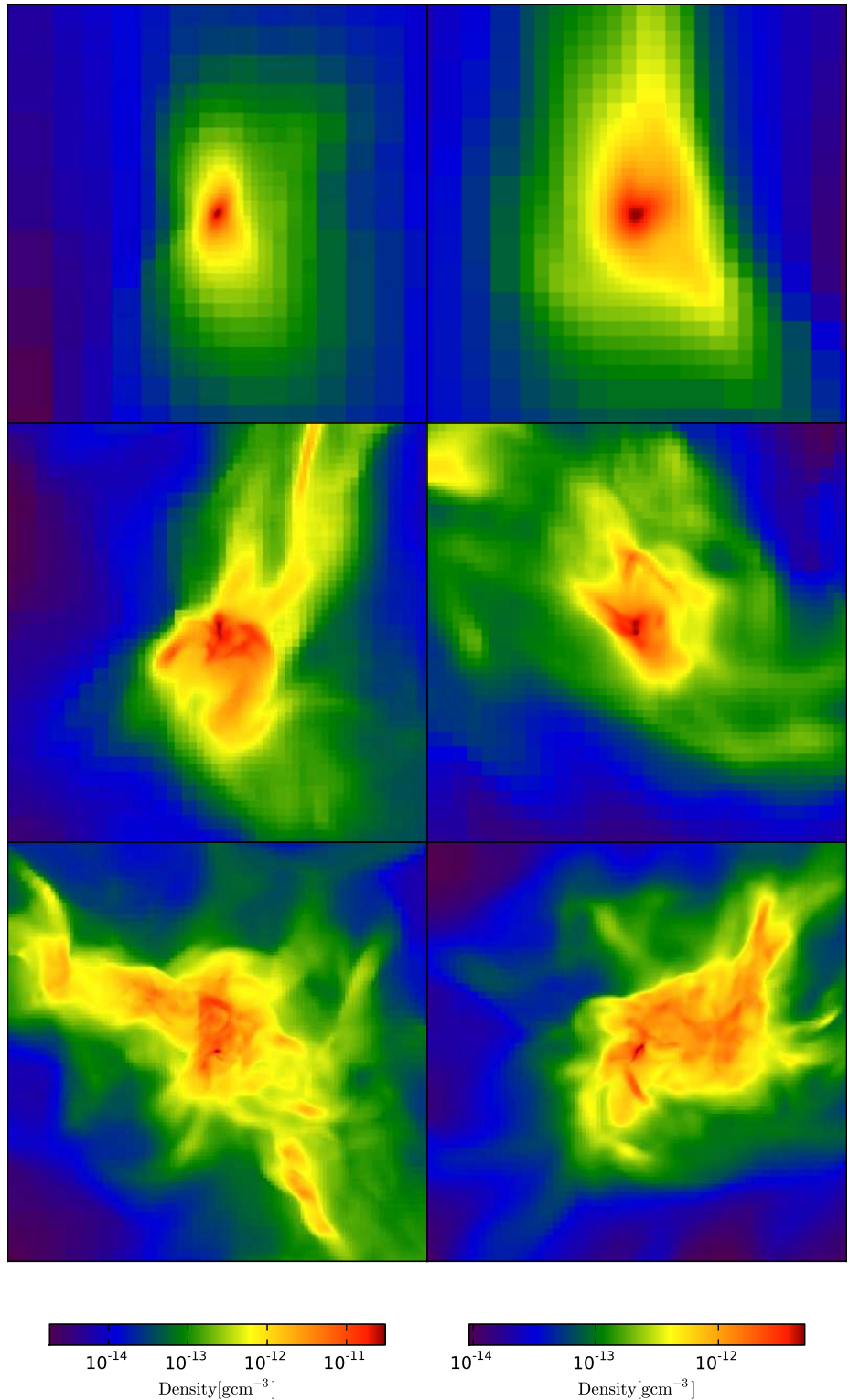}
\end{center}
\caption{The density distribution in the central $500$~AU with and without the turbulent subgrid model, and for different resolutions per Jeans length (top to bottom: $16$, $32$ and $64$ cells per Jeans length, left: standard hydro, right: hydro+subgrid-scale mode). We find that increasing the resolution per Jeans length considerably changes the morphology of the halo, as we start resolving turbulent structures at high resolution. The subgrid-scale model favors the formation of more compact structures and reduces the number of density peaks \citep{Latif12b}.}
\label{morph}
\end{figure*}

\begin{figure}
\includegraphics[scale=0.41]{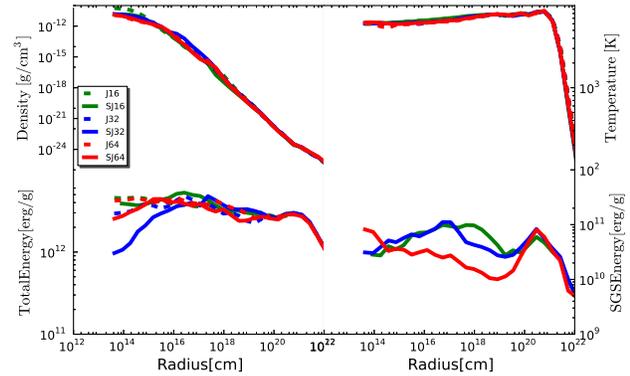}
\caption{The radial profiles of our halo at $z=11.9$ (top left: gas density, top right: gas temperature, bottom left: total energy, bottom right: subgrid-scale energy) for different resolutions per Jeans length, with and without the subgrid-scale model. Unlike the detailed morphology, we find that the radial profiles already converge for moderate resolutions per Jeans length \citep{Latif12b}.}
\label{prof}
\end{figure}

The halo collapses at a redshift of $11.9$ with a total mass of $8.06\times10^6$~M$_\odot$. The resulting density distributions in the central $500$~AU are given in Fig.~\ref{morph} for  different resolutions per Jeans length, comparing the results with and without the subgrid-scale model. While the low-resolution runs may give the impression of rather smooth density structures, a transition to turbulent, complex structures becomes visible for a resolution of $32$ cells per Jeans length, which is more pronounced when $64$ cells are adopted. This trend occurs both in the runs with and without the subgrid-scale model. The morphologies may however change considerably in the presence of the subgrid model, especially for the high-resolution results. The latter provides an indication that turbulent structures are not converged even in the highest resolution runs. For the radial profiles of the halo, we do however find convergence, as illustrated in Fig.~\ref{prof}. 

\begin{figure}
\includegraphics[scale=0.38]{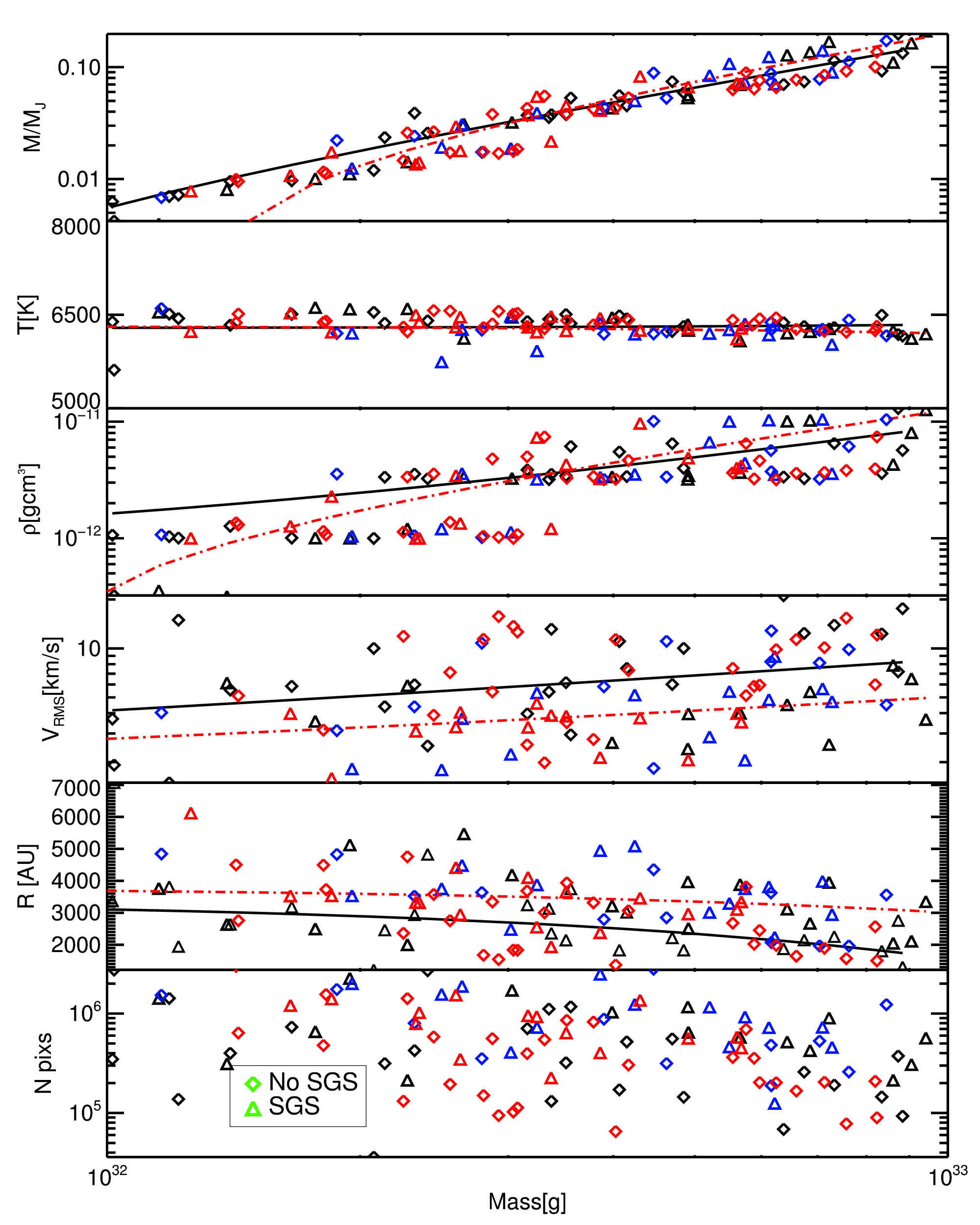}
\caption{The properties of gas clumps in the central $500$~AU for three different halos, calculated with and without the subgrid-scale model and for a resolution of $64$ cells per Jeans length. At this stage, the clumps are still unbound, and many of their properties follow characteristic power-laws. We note that the number of low-mass clumps is strongly reduced in the presence of the subgrid-scale model \citep{Latif12b}. Colors correspond to different halos, triangles denote simulations with, diamonds without the subgrid-scale model.  }
\label{clumps}
\end{figure}

We have examined the trends described here for a set of three massive halos, as described in more detail by \citet{Latif12b}. For these halos, we calculated the properties of gas clumps within the central $500$~AU using the clump finder of \citet{Williams94}. As shown in Fig.~\ref{clumps}, the latter exhibit a rather similar power-law behavior in all halos considered. We further find that the formation of low-mass clumps is suppressed in the presence of the subgrid-scale turbulence model.

\begin{table*}  
    \begin{tabular}{llccc}
      \parbox[0pt][2.5em][c]{0cm}{}  Model and reference     & label            & $\vartheta$         & $\bar{\Gamma}$                                                                                                                             ($\mathrm{Pm}\rightarrow\infty$)\\
      \hline
        \citet[][]{Kolmogorov41}		& K41	    								&  $1/3$              &  $\frac{37}{36}~\mathrm{Re}^{1/2}$ \\
      Intermittency of Kolmogorov turbulence  \citep[][]{She94}  & SL94 &  $0.35$  & $0.94~\mathrm{Re}^{0.48}$\\
       Driven supersonic MHD-turbulence \citep[][]{Boldyrev02}& BNP02 &  $0.37$    &  $0.84~\mathrm{Re}^{0.46}$   \\
       Observation in molecular clouds  \citep{Larson81}  & L81 &  $0.38$     &  $0.79~\mathrm{Re}^{0.45}$     \\
       Solenoidal forcing of the turbulence \citep{Federrath10turb} &FRKSM10 &  $0.43$  &                                                                                                                                  $0.54~\mathrm{Re}^{0.40}$  \\
       Compressive forcing of the turbulence \citep{Federrath10turb} &  FRKSM10 & $0.47$     &                                                                                                                              $0.34~\mathrm{Re}^{0.36}$  \\                     
             Observations in molecular clouds \citep{Ossenkopf02}    & OM02 &  $0.47$       &     $0.34~\mathrm{Re}^{0.36}$        \\ 
       \citet{Burgers48}                             & B48 & $1/2$      &  $\frac{11}{60}~\mathrm{Re}^{1/3}$ \\
    \end{tabular}
  \caption{The normalized growth rate of the small-scale dynamo $\bar{\Gamma}$ in the limit of infinite magnetic Prandtl number \citep{Schober12c}. We show our results for different types of turbulence, which are characterized by the exponent $\vartheta$ of the slope of the turbulent velocity spectrum, $v(\ell)\propto\ell^\vartheta$. The extreme values of $\vartheta$ are $1/3$ for Kolmogorov turbulence and $1/2$ for Burgers turbulence.}
  \label{ResultsTable}
\end{table*}

\section{Turbulent amplification of magnetic fields}
The turbulence present in these halos may efficiently amplifiy even weak  magnetic fields via the small-scale dynamo. Appropriate seeds may result from the Biermann battery \citep{Biermann50}, the Weibel instability \citep{Medvedev04, Lazar09}, thermal plasma fluctuations \citep{Schlickeiser12} or even the pre-recombination Universe \citep{Grasso01, Banerjee03}.

The small-scale dynamo has been explored with numerical simulations and analytical models \citep{Brandenburg05}. An analytical treatment is possible in terms of the Kazantsev model \citep{Kazantsev68, Subramanian98, Boldyrev04, Schober12c}, describing the dynamo amplification as the advection of passive vectors in a velocity field consisting of homogeneous and isotropic turbulence. The model describes magnetic field amplification in the kinematic regime, before backreactions occur. We note that two phases are relevant for the origin of high-redshift magnetic fields: While the kinematic phase leads to a fast exponential growth on timescales comparable to the eddy-turnover time in the viscous range, the non-linear growth leads to the formation of larger-scale magnetic fields. The overall process is expected to saturate within a few eddy-turnover times.

\subsection{The kinematic regime}
We start our considerations with the kinematic regime of dynamo amplification. The latter is described in terms of the induction equation, which is given as
\begin{equation}\label{eq:induction}
	\frac{\partial {\bf B}}{\partial t} = \nabla\times {\bf v}\times {\bf B} - \eta\nabla\times\nabla\times {\bf B},
\end{equation}
where {\bf B} is the magnetic field,  {\bf v} is the fluid velocity and $\eta$ the magnetic diffusivity. In the absence of helicity, the correlation function of the magnetic fields, $M_{ij}(r)=\langle B_i(0)B_j(r)\rangle$, can be decomposed into a transversal and longitudinal component:
\begin{equation}\label{eq:correlation}
	M_{ij}(r) = \left(\delta_{ij}-\frac{r_ir_j}{r^2}\right)M_N(r) + \frac{r_ir_j}{r^2}M_L(r).
\end{equation}
As a result of the constraint equation, $\nabla\cdot{\bf B}=0$, it can be shown that\begin{equation}
M_N=\frac{1}{2r}\frac{\partial }{\partial r}\left(r^2 M_L(r)\right).
\end{equation}
The same decomposition can be performed for the turbulence correlation function $T_{ij}(r)=\langle v_i(0)v_j(r) \rangle$. As described by \citet{Schober12c}, we adopt the following parametrizations for the longitudinal  correlation function,
\begin{equation}
  T_\text{L}(r) = \begin{cases}
               c\left(1-\mathrm{Re}^{(1-\vartheta)/(1+\vartheta)}\left(\frac{r}{L}\right)^{2}\right) & 0<r<\ell_\nu \\
               c\left(1-\left(\frac{r}{L}\right)^{\vartheta+1}\right)                   & \ell_\nu<r<L \\
               0                                                                               &  L<r,
            \end{cases}
            \end{equation}
           and the transversal correlation function,\begin{equation}
  T_\text{N}(r) = \begin{cases}
                     c\left(1-t_\vartheta \mathrm{Re}^{(1-\vartheta)/(1+\vartheta)} \left(\frac{r}{L}\right)^{2}\right)      & 0<r<\ell_\nu \\
               c\left(1-t_\vartheta\left(\frac{r}{L}\right)^{\vartheta+1}\right)                              & \ell_\nu<r<L \\
               0                                                                                                       & L<r,
                \end{cases}
\end{equation}
with $c=VL/3$ and $t_\vartheta=(21-38\vartheta)/5$. The turbulent slope is given as $\vartheta=1/3$ for Kolmogorov and $\vartheta=1/2$ for Burgers. The results for a larger set of models were derived by  \citet{Schober12c} and are given in Table~\ref{ResultsTable}. 

The latter are obtained employing an ansatz\begin{equation}
M_L(r,t)=~\Psi(r)e^{2\Gamma t}/(r^2\sqrt\kappa(r)),
\end{equation}
where $t$ denotes the time coordinate, $\Gamma$ can be interpreted as the growth rate of the correlation function, and $\kappa(r)$ a function describing the turbulent diffusion. From this ansatz, it is possible to obtain the so-called Kazantsev equation,
\begin{equation}\label{eq:radial}
	-\kappa(r)\frac{d^2\Psi(r)}{dr^2} + U(r)\Psi(r) = -\Gamma\Psi(r),
\end{equation}
which has the same form as the time-independent quantum-mechanical Schr\"odinger equation with a potential $U(r)$ given as
\begin{equation}\label{eq:potential}
	U(r) = \frac{\kappa''}{2} -\frac{(\kappa')^2}{4\kappa} + \frac{2T'_N}{r} + \frac{2(T_L - T_N - \kappa)}{r^2}.
\end{equation}

It can thus be solved using quantum-mechanical methods, in particular the so-called WKB-approximation. Introducing the normalized growth rate $\bar{\Gamma}$ as
\begin{equation}
	\bar{\Gamma} = \frac{L}{V}\Gamma, 
\end{equation}

\begin{figure}[h]
\includegraphics[scale=0.6]{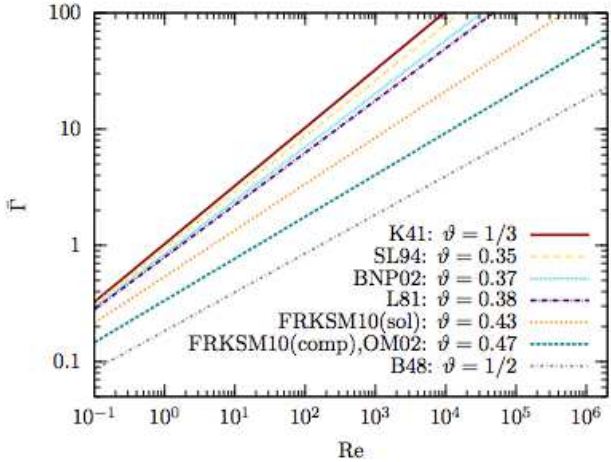}
\caption{The normalized growth rates of the small-scale dynamo for different turbulence models as a function of the Reynols number \citep{Schober12c}. The labels are defined in Table~\ref{ResultsTable}.}
\label{growth}
\end{figure}

\citet{Schober12c} have derived the resulting amplification rates in the limit of high magnetic Prandtl numbers Pm$\ \gg1$, defined as the ratio of kinetic and magnetic diffusivity. This limiting case is indeed appropriate for the interstellar and intergalactic medium, while stars and planets typically have magnetic Prandtl numbers of Pm$\ \ll1$. The resulting normalized growth rates are given in Table~\ref{ResultsTable}. The normalized growth rate is plotted as a function of the Reynolds number in Fig.~\ref{growth}. We find that the type of turbulence, and in particular its spectral slope has a major influence on the resulting amplification rate, which can be reduced by a few orders of magnitude. Nevertheless, we note that these growth rates are still orders of magnitudes smaller than the dynamical timescale of the system, implying that magnetic field amplification may still proceed in a highly efficient fashion in any of these cases \citep[e.g.][]{FederrathPRL}.

\subsection{The non-linear regime}
After saturation has occured on the viscous scale, where amplification originally is fastest, the magnetic field will be strong enough  to prevent further growth on that scale. However, as described by \citet{Scheko02}, magnetic field amplification will continue on larger scales $\ell$, with the typical eddy-turnover time $v_\ell \propto \ell^\vartheta$ on that scale. For Kolmogorov turbulence, the latter leads to a phase of linear growth of the magnetic energy, and one can show that Burgers turbulence leads to a quadratic increase of magnetic energy with time. During that phase, the length scale of the peak magnetic fluctuations increases considerably, from the viscous scale up to the injection scale of turbulence. At that point, the system is saturated. At low to moderate Mach numbers, the magnetic energy consists of $30-60\%$ of the turbulent energy, while at high Mach numbers, smaller saturation values of $\sim5\%$ are possible \citep{FederrathPRL}.

The time estimates required to reach saturation in the non-linear regime range from a few to a few dozen eddy-turnover times \citep{Scheko02, Beresnyak12}.  As a result, strong magnetic fields can be expected shortly after the formation of the first virialized objects due to turbulent amplification.

\subsection{Magnetic fields during primordial star formation}
\begin{figure}
\includegraphics[scale=0.76]{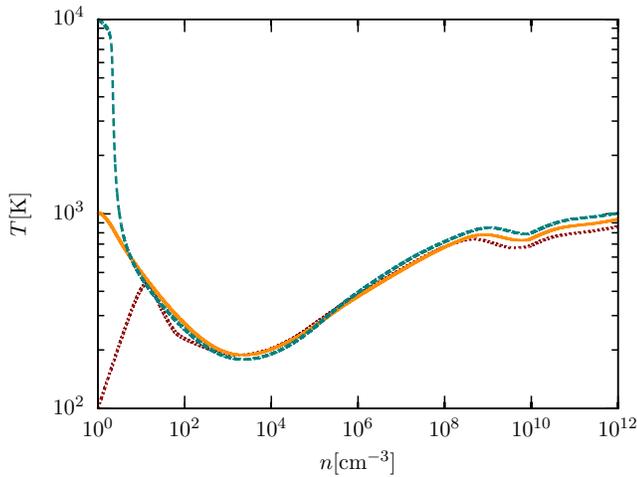}
\caption{The thermal evolution in primordial gas during gravitational collapse, as calculated by \citet{Schober12} with the one-zone model of \citet{Glover09}.}
\label{therm}
\end{figure}

In order to explore the consequences of turbulent magnetic field amplification for primordial star formation, \citet{Schober12} have implemented a magnetic field amplification model within the one-zone framework for primordial collapse developed by \citet{Glover09}. Their model assumes gravitational collapse on the free-fall timescale, solves the rate equations for primordial species along with the cooling/heating balance for the gas temperature (see Fig.~\ref{therm} for the thermal evolution). For the initial magnetic field, we adopted a typical value of $10^{-20}$~G, corresponding to the magnetic field strength from a Biermann battery. For the initial growth of the field during the kinematic regime, we employ the growth rates derived by \citet{Schober12c}. In this calculation, we assumed that turbulence is injected on the Jeans scale as a result of gravitational collapse \citep{Klessen10, Federrath11}, with typical turbulent velocities corresponding to the typical sound speed of the gas. After saturation occurs on the viscous scale, the evolution of the magnetic field is calculated employing the model of \citet{Scheko02}. We assume that the magnetic field is saturated when equipartition with turbulent energy is reached.

\begin{figure}
\includegraphics[scale=0.76]{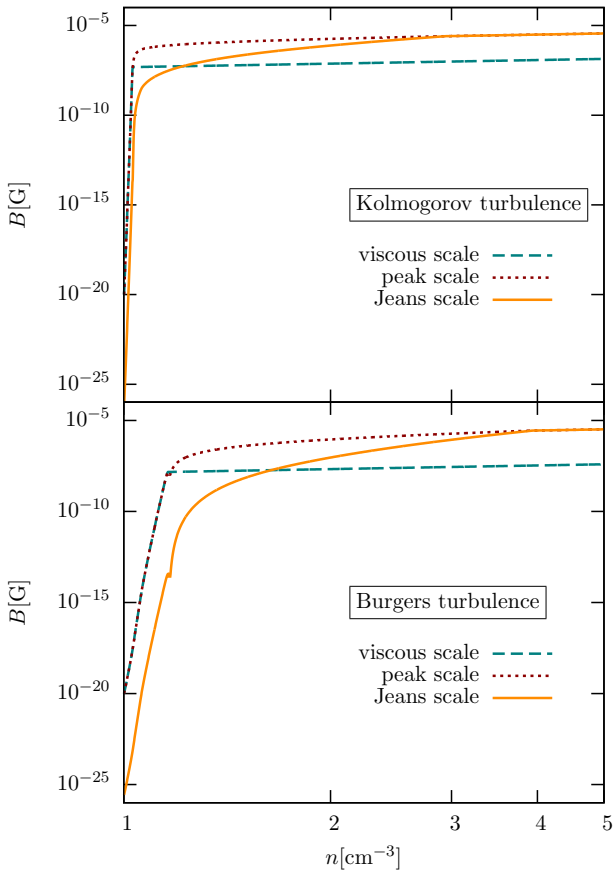}
\caption{The growth of magnetic fields during gravitational collapse in primordial minihalos, as calculated by \citet{Schober12}, for Kolmogorov and Burgers turbulence.}
\label{collB}
\end{figure}

We find that the magnetic field strength rises quickly during the kinematic regime, as the characteristic time scales are considerably smaller than the dynamical time. In the non-linear regime, the growth rate decreases when the saturation level is approached, which occurs after only a modest increase in density. The resulting field strength at densities of a few atoms per cm$^3$ is thus of the order of $10^{-5}$~G. The evolution of the field strength both for Kolmogorov and Burgers type turbulence is illustrated in Fig.~\ref{collB}. 

Overall, we  conclude that magnetic field amplification is highly efficient in the early Universe, resulting from the ubiquity of turbulence and the efficient amplification via the small-scale dynamo. We expect that such turbulent field structures can be probed in high-redshift starbursts with the Square Kilometre Array\footnote{SKA website: http://www.ska.ac.za/}.

\acknowledgements

DRGS, JS, RSK and SB acknowledge funding from the {\em Deutsche Forschungsgemeinschaft} (DFG) in the {\em Schwerpunktprogramm} SPP 1573 ``Physics of the Interstellar Medium" under grants KL 1358/14-1 and SCHL 1964/1-1. DRGS, JN and WS~thank for funding via the SFB 963/1 on ``Astrophysical flow instabilities and turbulence". JS acknowledges the support by IMPRS HD, the HGSFP and the SFB~881 ''The Milky Way System''. CF thanks for an ARC Discovery Projects Fellowship (DP110102191).


\begin{thebibliography}{52}
\expandafter\ifx\csname natexlab\endcsname\relax\def\natexlab#1{#1}\fi

\bibitem[{{Abel} {et~al.}(2002){Abel}, {Bryan}, \& {Norman}}]{Abel02}
{Abel}, T., {Bryan}, G.~L., \& {Norman}, M.~L. 2002, Science, 295, 93

\bibitem[{{Banerjee} \& {Jedamzik}(2003)}]{Banerjee03}
{Banerjee}, R., \& {Jedamzik}, K. 2003, Physical Review Letters, 91, 251301

\bibitem[{{Beck}(2004)}]{Beck04}
{Beck}, R. 2004, APS\&S, 289, 293

\bibitem[{{Beresnyak}(2012)}]{Beresnyak12}
{Beresnyak}, A. 2012, Physical Review Letters, 108, 035002

\bibitem[{{Bernet} {et~al.}(2008){Bernet}, {Miniati}, {Lilly}, {Kronberg}, \&
  {Dessauges-Zavadsky}}]{Bernet08}
{Bernet}, M.~L., {Miniati}, F., {Lilly}, S.~J., {Kronberg}, P.~P., \&
  {Dessauges-Zavadsky}, M. 2008, Nature, 454, 302

\bibitem[{{Biermann}(1950)}]{Biermann50}
{Biermann}, L. 1950, Zeitschrift Naturforschung Teil A, 5, 65

\bibitem[{{Boldyrev} \& {Cattaneo}(2004)}]{Boldyrev04}
{Boldyrev}, S., \& {Cattaneo}, F. 2004, Physical Review Letters, 92, 144501

\bibitem[{{Boldyrev} {et~al.}(2002){Boldyrev}, {Nordlund}, \&
  {Padoan}}]{Boldyrev02}
{Boldyrev}, S., {Nordlund}, {\AA}., \& {Padoan}, P. 2002, Physical Review
  Letters, 89, 031102

\bibitem[{{Brandenburg} \& {Subramanian}(2005)}]{Brandenburg05}
{Brandenburg}, A., \& {Subramanian}, K. 2005, Phys. Rep., 417, 1

\bibitem[{{Broderick} {et~al.}(2012){Broderick}, {Chang}, \&
  {Pfrommer}}]{Broderick12}
{Broderick}, A.~E., {Chang}, P., \& {Pfrommer}, C. 2012, ApJ, 752, 22

\bibitem[{Burgers(1948)}]{Burgers48}
Burgers, J. 1948, Advances in Applied Mechanics, Vol.~1, {A Mathematical Model
  Illustrating the Theory of Turbulence} (Elsevier), 171 -- 199

\bibitem[{{Chy{\.z}y} {et~al.}(2011){Chy{\.z}y}, {We{\.z}gowiec}, {Beck}, \&
  {Bomans}}]{Chyzy11}
{Chy{\.z}y}, K.~T., {We{\.z}gowiec}, M., {Beck}, R., \& {Bomans}, D.~J. 2011,
  A\&A, 529, A94

\bibitem[{{Federrath} {et~al.}(2011{\natexlab{a}}){Federrath}, {Chabrier},
  {Schober}, {Banerjee}, {Klessen}, \& {Schleicher}}]{FederrathPRL}
{Federrath}, C., {Chabrier}, G., {Schober}, J., {Banerjee}, R., {Klessen},
  R.~S., \& {Schleicher}, D.~R.~G. 2011{\natexlab{a}}, Physical Review Letters,
  107, 114504

\bibitem[{{Federrath} {et~al.}(2010){Federrath}, {Roman-Duval}, {Klessen},
  {Schmidt}, \& {Mac Low}}]{Federrath10turb}
{Federrath}, C., {Roman-Duval}, J., {Klessen}, R.~S., {Schmidt}, W., \& {Mac
  Low}, M.-M. 2010, A\&A, 512, A81

\bibitem[{{Federrath} {et~al.}(2011{\natexlab{b}}){Federrath}, {Sur},
  {Schleicher}, {Banerjee}, \& {Klessen}}]{Federrath11}
{Federrath}, C., {Sur}, S., {Schleicher}, D.~R.~G., {Banerjee}, R., \&
  {Klessen}, R.~S. 2011{\natexlab{b}}, ApJ, 731, 62

\bibitem[{{Glover} \& {Savin}(2009)}]{Glover09}
{Glover}, S.~C.~O., \& {Savin}, D.~W. 2009, MNRAS, 393, 911

\bibitem[{{Grasso} \& {Rubinstein}(2001)}]{Grasso01}
{Grasso}, D., \& {Rubinstein}, H.~R. 2001, Physics Reports, 348, 163

\bibitem[{{Greif} {et~al.}(2008){Greif}, {Johnson}, {Klessen}, \&
  {Bromm}}]{Greif08}
{Greif}, T.~H., {Johnson}, J.~L., {Klessen}, R.~S., \& {Bromm}, V. 2008,
  \mnras, 387, 1021

\bibitem[{{Heesen} {et~al.}(2011){Heesen}, {Beck}, {Krause}, \&
  {Dettmar}}]{Heesen11}
{Heesen}, V., {Beck}, R., {Krause}, M., \& {Dettmar}, R.-J. 2011, A\&A, 535,
  A79

\bibitem[{{Kazantsev}(1968)}]{Kazantsev68}
{Kazantsev}, A.~P. 1968, Sov. Phys. JETP, 26, 1031

\bibitem[{{Kepley} {et~al.}(2011){Kepley}, {Zweibel}, {Wilcots}, {Johnson}, \&
  {Robishaw}}]{Kepley11}
{Kepley}, A.~A., {Zweibel}, E.~G., {Wilcots}, E.~M., {Johnson}, K.~E., \&
  {Robishaw}, T. 2011, ApJ, 736, 139

\bibitem[{{Kim} {et~al.}(1990){Kim}, {Kronberg}, {Dewdney}, \&
  {Landecker}}]{Kim90}
{Kim}, K.-T., {Kronberg}, P.~P., {Dewdney}, P.~E., \& {Landecker}, T.~L. 1990,
  \apj, 355, 29

\bibitem[{{Klessen} \& {Hennebelle}(2010)}]{Klessen10}
{Klessen}, R.~S., \& {Hennebelle}, P. 2010, A\&A, 520, A17+

\bibitem[{{Kolmogorov}(1941)}]{Kolmogorov41}
{Kolmogorov}, A. 1941, Akademiia Nauk SSSR Doklady, 30, 301

\bibitem[{{Kronberg} {et~al.}(2008){Kronberg}, {Bernet}, {Miniati}, {Lilly},
  {Short}, \& {Higdon}}]{Kronberg08}
{Kronberg}, P.~P., {Bernet}, M.~L., {Miniati}, F., {Lilly}, S.~J., {Short},
  M.~B., \& {Higdon}, D.~M. 2008, \apj, 676, 70

\bibitem[{{Larson}(1981)}]{Larson81}
{Larson}, R.~B. 1981, MNRAS, 194, 809

\bibitem[{{Latif} {et~al.}(2012){Latif}, {Schleicher}, {Schmidt}, \&
  {Niemeyer}}]{Latif12b}
{Latif}, M.~A., {Schleicher}, D.~R.~G., {Schmidt}, W., \& {Niemeyer}, J. 2012,
  MNRAS, submitted (ArXiv e-prints 1210.1802)

\bibitem[{{Lazar} {et~al.}(2009){Lazar}, {Schlickeiser}, {Wielebinski}, \&
  {Poedts}}]{Lazar09}
{Lazar}, M., {Schlickeiser}, R., {Wielebinski}, R., \& {Poedts}, S. 2009, \apj,
  693, 1133

\bibitem[{{Maier} {et~al.}(2009){Maier}, {Iapichino}, {Schmidt}, \&
  {Niemeyer}}]{Maier09}
{Maier}, A., {Iapichino}, L., {Schmidt}, W., \& {Niemeyer}, J.~C. 2009, \apj,
  707, 40

\bibitem[{{Medvedev} {et~al.}(2004){Medvedev}, {Silva}, {Fiore}, {Fonseca}, \&
  {Mori}}]{Medvedev04}
{Medvedev}, M.~V., {Silva}, L.~O., {Fiore}, M., {Fonseca}, R.~A., \& {Mori},
  W.~B. 2004, Journal of Korean Astronomical Society, 37, 533

\bibitem[{{Miniati} \& {Elyiv}(2012)}]{Miniati12}
{Miniati}, F., \& {Elyiv}, A. 2012, ArXiv e-prints

\bibitem[{{Murphy}(2009)}]{Murphy09}
{Murphy}, E.~J. 2009, \apj, 706, 482

\bibitem[{{Neronov} \& {Vovk}(2010)}]{Neronov10}
{Neronov}, A., \& {Vovk}, I. 2010, Science, 328, 73

\bibitem[{{O'Shea} {et~al.}(2004){O'Shea}, {Bryan}, {Bordner}, {Norman},
  {Abel}, {Harkness}, \& {Kritsuk}}]{OShea04}
{O'Shea}, B.~W., {Bryan}, G., {Bordner}, J., {Norman}, M.~L., {Abel}, T.,
  {Harkness}, R., \& {Kritsuk}, A. 2004, ArXiv Astrophysics e-prints

\bibitem[{{Ossenkopf} \& {Mac Low}(2002)}]{Ossenkopf02}
{Ossenkopf}, V., \& {Mac Low}, M.-M. 2002, A\&A, 390, 307

\bibitem[{{Peters} {et~al.}(2012){Peters}, {Schleicher}, {Klessen}, {Banerjee},
  {Federrath}, {Smith}, \& {Sur}}]{Peters12}
{Peters}, T., {Schleicher}, D.~R.~G., {Klessen}, R.~S., {Banerjee}, R.,
  {Federrath}, C., {Smith}, R.~J., \& {Sur}, S. 2012, ArXiv e-prints 1209.5861

\bibitem[{{Schekochihin} {et~al.}(2002){Schekochihin}, {Cowley}, {Hammett},
  {Maron}, \& {McWilliams}}]{Scheko02}
{Schekochihin}, A.~A., {Cowley}, S.~C., {Hammett}, G.~W., {Maron}, J.~L., \&
  {McWilliams}, J.~C. 2002, New Journal of Physics, 4, 84

\bibitem[{{Schleicher} {et~al.}(2010){Schleicher}, {Banerjee}, {Sur},
  {Arshakian}, {Klessen}, {Beck}, \& {Spaans}}]{Schleicher10c}
{Schleicher}, D.~R.~G., {Banerjee}, R., {Sur}, S., {Arshakian}, T.~G.,
  {Klessen}, R.~S., {Beck}, R., \& {Spaans}, M. 2010, A\&A, 522, A115

\bibitem[{{Schlickeiser}(2012)}]{Schlickeiser12}
{Schlickeiser}, R. 2012, ArXiv e-prints 1207.2963

\bibitem[{{Schmidt} \& {Federrath}(2011)}]{Schmidt11}
{Schmidt}, W., \& {Federrath}, C. 2011, A\&A, 528, A106

\bibitem[{{Schober} {et~al.}(2012{\natexlab{a}}){Schober}, {Schleicher},
  {Federrath}, {Glover}, {Klessen}, \& {Banerjee}}]{Schober12}
{Schober}, J., {Schleicher}, D., {Federrath}, C., {Glover}, S., {Klessen},
  R.~S., \& {Banerjee}, R. 2012{\natexlab{a}}, \apj, 754, 99

\bibitem[{{Schober} {et~al.}(2012{\natexlab{b}}){Schober}, {Schleicher},
  {Federrath}, {Klessen}, \& {Banerjee}}]{Schober12c}
{Schober}, J., {Schleicher}, D., {Federrath}, C., {Klessen}, R., \& {Banerjee},
  R. 2012{\natexlab{b}}, PRE, 85, 026303

\bibitem[{{She} \& {Leveque}(1994)}]{She94}
{She}, Z.-S., \& {Leveque}, E. 1994, Physical Review Letters, 72, 336

\bibitem[{{Subramanian}(1998)}]{Subramanian98}
{Subramanian}, K. 1998, MNRAS, 294, 718

\bibitem[{{Sur} {et~al.}(2012){Sur}, {Federrath}, {Schleicher}, {Banerjee}, \&
  {Klessen}}]{Sur12}
{Sur}, S., {Federrath}, C., {Schleicher}, D.~R.~G., {Banerjee}, R., \&
  {Klessen}, R.~S. 2012, MNRAS, 423, 3148

\bibitem[{{Sur} {et~al.}(2010){Sur}, {Schleicher}, {Banerjee}, {Federrath}, \&
  {Klessen}}]{Sur10}
{Sur}, S., {Schleicher}, D.~R.~G., {Banerjee}, R., {Federrath}, C., \&
  {Klessen}, R.~S. 2010, ApJL, 721, L134

\bibitem[{{Takahashi} {et~al.}(2012){Takahashi}, {Mori}, {Ichiki}, \&
  {Inoue}}]{Takahashi12}
{Takahashi}, K., {Mori}, M., {Ichiki}, K., \& {Inoue}, S. 2012, ApJL, 744, L7

\bibitem[{{Tavecchio} {et~al.}(2011){Tavecchio}, {Ghisellini}, {Bonnoli}, \&
  {Foschini}}]{Tavecchio11}
{Tavecchio}, F., {Ghisellini}, G., {Bonnoli}, G., \& {Foschini}, L. 2011,
  MNRAS, 414, 3566

\bibitem[{{Turk} {et~al.}(2012){Turk}, {Oishi}, {Abel}, \& {Bryan}}]{Turk12}
{Turk}, M.~J., {Oishi}, J.~S., {Abel}, T., \& {Bryan}, G.~L. 2012, ApJ, 745,
  154

\bibitem[{{Williams} {et~al.}(1994){Williams}, {de Geus}, \&
  {Blitz}}]{Williams94}
{Williams}, J.~P., {de Geus}, E.~J., \& {Blitz}, L. 1994, \apj, 428, 693

\bibitem[{{Wise} \& {Abel}(2008)}]{Wise08}
{Wise}, J.~H., \& {Abel}, T. 2008, ApJ, 685, 40

\bibitem[{{Y{\"u}ksel} {et~al.}(2012){Y{\"u}ksel}, {Stanev}, {Kistler}, \&
  {Kronberg}}]{Yuksel12}
{Y{\"u}ksel}, H., {Stanev}, T., {Kistler}, M.~D., \& {Kronberg}, P.~P. 2012,
  \apj, 758, 16

\end{thebibliography}

\end{document}